\documentclass[aps,pra,twocolumn,showpacs]{revtex4}
\usepackage{epsfig}

\newcommand{\p}{\ensuremath{\hat p}}
\newcommand{\q}{\ensuremath{\hat q}}

\renewcommand{\a}{\ensuremath{\hat a}}
\newcommand{\ave}[1]{\ensuremath{\langle #1 \rangle}}
\newcommand{\D}{\Delta}
\newcommand{\de}{\delta}
\newcommand{\e}{\mathrm{e}}
\newcommand{\A}{\alpha}
\newcommand{\g}{\gamma}
\newcommand{\be}{\begin{equation}}
\newcommand{\ee}{\end{equation}}
\newcommand{\ba}{\begin{eqnarray}}
\newcommand{\ea}{\end{eqnarray}}
\def\lsim{\mathrel{\rlap{\lower4pt\hbox{\hskip0pt$\sim$}}
    \raise1pt\hbox{$<$}}}                
\def\gsim{\mathrel{\rlap{\lower4pt\hbox{\hskip0pt$\sim$}}
    \raise1pt\hbox{$>$}}} 
\DeclareMathSizes{10}{10}{5}{3}

\begin{document}

\title{Scalable continuous-variable entanglement of light beams produced by optical parametric oscillators}
\author{Katiuscia N. Cassemiro$^1$\email{nadyne@if.usp.br} and Alessandro S. Villar$^{2,3}$}
\affiliation{
\mbox{$^1$ Instituto de F\'\i sica, Universidade de S\~ao Paulo, Caixa Postal 66318, 05315-970 S\~ao Paulo, SP, Brazil}\\
\mbox{$^2$ Institut f\"ur Experimentalphysik, Universit\"at Innsbruck, Technikerstr. 25/4, 6020 Innsbruck, Austria} \\
\mbox{$^3$ Institute for Quantum Optics and Quantum Information, Austrian Academy of Sciences, 6020 Innsbruck, Austria}} 

\begin{abstract}
We show that scalable multipartite entanglement among light fields may be generated by optical parametric oscillators (OPO). The tripartite entanglement existent among the three bright beams produced by a single OPO -- pump, signal, and idler -- is scalable to a system of many OPOs by pumping them in cascade with the same optical field. This latter serves as an entanglement distributor. The special case of two OPOs is studied, as it is shown that the resulting five bright beams share genuine multipartite entanglement. In addition, the structure of entanglement distribution among the fields can be manipulated to some degree by tuning the incident pump power. The scalability to many fields is straightforward, allowing an alternative implementation of a multipartite quantum information network with continuous variables. 
\end{abstract}


\maketitle


\section{Introduction}
\label{secintro}

Entanglement is the primary resource in the field of quantum information. Its strikingly singular properties, which permeated the early debates on the fundamentals of quantum mechanics~\cite{epr_physrev35}, are nowadays employed to take computation technology and communication security beyond the classical limit~\cite{bennettvicenzoreview,ekertprl91}. Some applications have been successfully demonstrated in systems composed of up to four parties~\cite{telet-tripa-furusawa,pengentswap}, but their unrestricted use in quantum computation and networks relies on the ability to entangle numerous physical systems~\cite{nielsen-chuang}. 

Multipartite entanglement brings a new level of complexity, and therefore novel possibilities, such as stronger violation of Bell-type inequalities~\cite{merminprl90}. The geometry of entanglement distribution among parties can be manipulated to perform one-way quantum computation~\cite{onewaycvcomput_prl01,cvclusterstatecomput_prl06} or to guarantee that confidential information is correctly addressed and safely transmitted~\cite{lanceprl04,weinfurterprl07}. When it comes to continuous-variable (CV) entanglement~\cite{ppt-review}, the quadratures of light fields -- equivalent to the intensity and phase in the case of bright beams -- offer the possibility to implement such ideas. 

The usual way to produce multipartite CV entangled light beams relies on the interference of at least one quantum field, presenting quadrature squeezing, with any desired number of vacuum fields using beam splitters~\cite{telet-tripa-furusawa,multiteleportnetw}. This has been realized for up to four squeezed beams by employing optical parametric amplifiers~\cite{clusterpeng}. However, the requirement for interference limits this type of scheme to fields with the same optical frequency. 

We present an alternative scalable way to directly generate CV multipartite entangled beams without the need for interference~\cite{pfisterpra04,chinapra07}. Our scheme utilizes as a basic resource the tripartite entanglement existent between the three fields -- signal, idler, and pump -- produced by an optical parametric oscillator (OPO) operating above threshold~\cite{prltrientangopo}. Various OPOs are used in a chain configuration, such that the pump beam reflected by the first one, already entangled to the first pair of down-converted twin beams, pumps a second OPO, entangling both pairs of twins, and so on. The pump beam then acts as a sequential entanglement distributor. Furthermore, we show that the entanglement structure can be manipulated by tuning the pump power. We study the simplest case of two OPOs to show that the resulting five light beams present genuine pentapartite entanglement. 

This paper is organized as follows. We begin in Sec.~\ref{secmulti} with the method we use to characterize multipartite entanglement, the positivity of the partially transposed (PPT) density matrix~\cite{ppt,simon}. We proceed in Sec.~\ref{secdescription} and~\ref{sectripa} with a brief review of the equations describing the quantum properties of the light beams generated by a single OPO. By using this result, it is straightforward to derive the set of equations for the pentapartite system. In Sec.~\ref{secscalable}, we present our results by applying the PPT criterion to the composite system of two OPOs, thus demonstrating the existence of genuine pentapartite entanglement among the light beams, and analyze the structure of entanglement distribution. Concluding remarks are presented in Sec.~\ref{secconclusion}.


\section{Multipartite Entanglement of Gaussian States}
\label{secmulti}

Gaussian states are completely characterized by their second order moments, organized in the covariance matrix $V=\ave{\vec \mathbf{x}^T \, \vec \mathbf{x}}$, where
\be
\vec \mathbf{x} = (\p_{1},\; \q_{1}, \;\p_{2}, \;\q_{2},\, \dots , \;\p_N, \;\q_N )
\ee
is the vector of the amplitude [$\p_j=\exp(-i\varphi_j)\,\a_j+\exp(i\varphi_j)\,\a_j^\dag$] and phase [$\q_j=-i(\exp(-i\varphi_j)\,\a_j-\exp(i\varphi_j)\,\a_j^\dag)$] quadrature operators, chosen relative to a phase $\varphi_j$, and $N$ is the number of field modes. The operators $\a_j$ and $\a_j^\dag$ are the usual annihilation and creation operators for mode $j$. The canonical commutation relations can be written in the compact form $[\vec x, \vec x^T]=2i \Omega$, where
\be
\Omega=\bigoplus_{j=1}^{N}\,\omega, \quad \omega=\left(\begin{array}{cc} 0 & 1 \\ -1 & 0 \end{array}\right) \, .
\ee

A typical unitary physical transformation acting on the Hilbert space of the system, such as the ones associated with beam splitters and one-- or two--mode squeezers, preserves the gaussian character of the state. The corresponding transformation $S$ acting on the covariance matrix pertains to the real symplectic group $S\in$~Sp($2N$,$\mathcal{R}$), which preserves the commutation relations, $S\Omega S^T=\Omega$~\cite{symplectic-simon}. 

In order to represent a physical state, the covariance matrix must respect the Robertson-Schr\"odinger uncertainty principle~\cite{robertson-schroedinger}, 
\be
\label{uncertprinc}
V+i\Omega\geq0.
\ee
This condition, which is invariant under symplectic transformations, implies a constraint to the symplectic eigenvalues of the covariance matrix, 
\ba
\label{uncprinsympeigenpt}
\nu_k\geq1\;,\; & k\in{1,2,\dots, N}\, .
\ea
The symplectic eigenvalues can be computed as the square roots of the ordinary eigenvalues of $-(\Omega\,V)^2$. They are invariant under local operations and classical communication (LOCC)~\cite{adessopra04}. If the covariance matrix corresponds to a pure state, then one necessarily has $\nu_k=1$, $\forall k$.

The existence of entanglement among the field modes may be unveiled by applying separability tests to all possible partitions, and its structure, also by the reduced subsystems. If all partitions of a pure state are inseparable, the system is said to present genuine $N$-partite entanglement. One powerful testing method relies on the positivity of the partially transposed (PPT) density matrix~\cite{ppt}. This map is positive regarding all separable states, but may be negative to entangled states. It is equivalent, in phase space, to a local inversion of time for the transposed subsystems~\cite{simon}. Regarding the covariance matrix, the PT operation takes $\q_j$ in $-\q_j$ for the desired subset of modes. The failure of the resulting PT covariance matrix $\tilde V$ to comply with the uncertainty principle~(\ref{uncertprinc}) is a sufficient condition for the existence of entanglement between the transposed subset and the remaining subset~\cite{simon}. 

Therefore, Eq.~\ref{uncertprinc} must be satisfied by all the possible transpositions of covariance matrices representing completely separable states. Equivalently, the symplectic eigenvalues $\tilde \nu$ of $\tilde V$ must fulfill Eq.~\ref{uncprinsympeigenpt} in this case. The PPT criterion is both necessary and sufficient for pure or mixed states split in partitions $1\times (N-1)$~\cite{simon,werner}. Other partitions from systems with $N\geq2$ may possess bound entanglement, a non-distillable form of entanglement which is not revealed by partial transposition~\cite{werner}. Nevertheless, it is always sufficient.

It is worth mentioning that the smallest symplectic eigenvalue $\tilde{\nu}_{min}$ of the PT covariance matrix is useful not only to witness the entanglement but also to quantify it. In fact, the entanglement measure given by the logarithmic negativity~\cite{measure-negat} can be written as a decreasing function of $\tilde{\nu}_{min}$, for all ($M+N$)-mode bisymmetric Gaussian states~\cite{adessopra04}. Thus, a larger violation of Eq.~(\ref{uncprinsympeigenpt}) implies a larger amount of entanglement.


\section{Description of the OPO}
\label{secdescription}

The optical parametric oscillator consists of a nonlinear $\chi^{(2)}$ crystal disposed inside an optical cavity, in this manner coupling three modes of the electromagnetic field. It is driven by an incident laser, the {\it pump beam}, at frequency $\omega_0$. Following the usual terminology, the fields generated by downconversion are called ``signal'' and ``idler,'' with frequencies $\omega_1$ and $\omega_2$ respecting, by energy conservation, $\omega_0=\omega_1+\omega_2$. Above the oscillation threshold, signal and idler are bright light beams known to possess strong quantum correlations in both intensity~\cite{firsttwin87} and phase~\cite{prlentangtwinopo}, therefore called ``twin beams''.

The covariance matrix of a single OPO can be found in many references~\cite{fabreqopt90,optcomm04}. The standard treatment begins with the master equation for the density operator, which is then converted to a Fokker-Planck equation for a quasi-probability distribution (in the present case, the Wigner function), and finally to a set of Langevin equations for its complex arguments representing the classical field amplitudes $\A_j(t)$. The labels $j \in \{0,1,2\}$ correspond to pump, signal and idler, respectively.  The set of Langevin equations describing the quantum fluctuations $\de \A_j$ of the intracavity fields $\A_j(t)$ are:
\ba 
\tau\frac{d}{dt}\de\A_0 &=& -\g_0(1-i\D_0)\de\A_0  +\hspace{-1pt}\sqrt{2\g_0}\de\A_0^\mathrm{in} -\nonumber \\ 
\label{eqsflutuopo1} && \hspace{-35pt} -\frac{ p}{p_0}\g (1+i \D)\e^{-i\theta}(\e^{i\varphi_2}\de\A_1+\e^{i\varphi_1}\de\A_2) , \\ 
\tau\frac{d}{dt}\de\A_1 &= &-\g (1-i\D)\de\A_1 +\sqrt{2\g}\de v_1+ \nonumber\\
\label{eqsflutuopo2} && \hspace{-35pt} +\frac{\g (1-i \D)}{p_0}\e^{i\theta}(p\,\e^{-i\varphi_2}\de\A_0+p_0\e^{i\varphi_0}\de\A_2^*), \\ 
\tau\frac{d}{dt}\de\A_2 &\hspace{-5pt}= &\hspace{-5pt}-\g (1-i\D)\de\A_2 +\sqrt{2\g}\de v_2+ \nonumber\\
\label{eqsflutuopo3} && \hspace{-35pt} +\frac{\g (1-i \D)}{p_0}\e^{i\theta}(p\,\e^{-i\varphi_1}\de\A_0+p_0\e^{i\varphi_0}\de\A_1^*),
\ea
where a linearization procedure has been applied, $\A_j(t)\equiv \bar\A_j+\de\A_j(t)$, such that terms involving the product of fluctuations have been ignored. The mean complex amplitude $\bar\A_j=p_j\,\exp(i\varphi_j)$ reproduces the classical result for the mean intracavity field amplitude $p_j$ and phase $\varphi_j$~\cite{optcomm04} (in the above equations, $\theta=\varphi_1+\varphi_2-\varphi_0$). The latter is defined with respect to the incoming pump beam $\A_0^\mathrm{in}$, chosen real. The coefficient $\gamma_j$ is related to the coupling mirror transmission for the field $j$, $T_j=2\gamma_j\ll1$. Only intracavity losses originating from mirror transmissions are considered in our analysis (although the effects of spurious losses are briefly discussed in the final results). The detuning between the OPO cavity and the field $\bar\alpha_j$ is given by $\D_j$, where $\D_1=\D_2\equiv\D$, and the cavity roundtrip time is $\tau$. We assume equal mirror transmissions for signal and idler beams, i.e. $\gamma_1=\gamma_2\equiv\gamma$, resulting in the equality of their mean amplitudes, $p_1=p_2\equiv p$. The terms $\de v_j$ are the vacuum fluctuations associated to these couplings. For the intracavity pump mode, they are due to the quantum fluctuations of the input pump laser beam, $\de v_0\equiv\de\A_0^\mathrm{in}$; however, since we consider it as shot-noise limited, its fluctuations obey the same statistics as the vacuum.

To obtain from Eqs.~(\ref{eqsflutuopo1})--(\ref{eqsflutuopo3}) equations for the quadratures, we write
\be
\label{flutuquad}
\de\A_j(t)= \frac{e^{i\varphi_j}}{2}\,[\,\de p_j(t)+ i\,\de q_j(t)\,]\;,
\ee
where the phase references for the definition of the quadratures were chosen as the respective mean fields' phases. We further simplify the problem by assuming exact triple resonance, $\D_0=\D=0$, since detunings have the main effect of weakly coupling amplitude and phase quadratures, but do not change the overall physical behavior of the system~\cite{optcomm04}. The solution is readily obtained in frequency domain by considering the combined quadratures $p_\pm=(p_1\pm p_2)/\sqrt{2}$ and $q_\pm=(q_1\pm q_2)/\sqrt{2}$, for which the evolution equations decouple~\cite{fabreqopt90}. The output quadratures are finally determined using the input-output relations,
\be
\de p_s^\mathrm{out}(\Omega)= -\de v_{p_s}(\Omega) + \sqrt{2\g}\de p_s(\Omega),
\ee
where $s\in\{0,+,-\}$. A similar relation holds for the phase quadrature. The solution is
\ba
\label{psumout}
\de p_+^\mathrm{out} &=& \kappa_p\,\de p_0^\mathrm{in} +(2\g\xi_p-1)\,\de v_{p_+}  \;, \\
\label{qsumout}
\de q_+^\mathrm{out} &=& \kappa_q\,\de q_0^\mathrm{in} +(2\g\xi_q-1)\,\de v_{q_+}  \;, \\
\label{p0out}
\de p_0^\mathrm{out} &=& \vartheta_p\,\de p_0^\mathrm{in} -\kappa_p\,\de v_{p_+}  \;, \\
\label{q0out}
\de q_0^\mathrm{out} &=& \vartheta_q\,\de q_0^\mathrm{in} -\kappa_q\,\de v_{q_+}  \;, \\
\label{pdifout}
\de p_-^\mathrm{out} &=& -\frac{i\Omega'}{1+i\Omega'}\,\de v_{p_-} \; ,\\
\label{qdifout}
\de q_-^\mathrm{out} &=& -i\frac{1-i\Omega'}{\Omega'}\de v_{q_-} \;,
\ea
where
\ba
\label{defxi}
\xi_p &=& 2i\g \Omega' +\frac{2\g ^2\beta^2}{\g_0+2i\g \Omega'} \;,\\
\label{defzeta}
\xi_q &=&  2\g  +\xi_p \;,\\
\label{defkappap}
\kappa_{p,q} &=&  \frac{2\sqrt{2}\g \beta\sqrt{\g_0\g}}{\g_0+2i\g \Omega'}\,(\xi_{p,q})^{-1} \;,\\
\label{defvarthetap}
\vartheta_{p,q} &\hspace{-6pt}=&\hspace{-6pt}  -1 +\frac{2\g_0}{\g_0+2i\g \Omega'}\left(1\hspace{-1pt}-\hspace{-3pt} \sqrt{2\g_o}\g^{\frac{3}{2}} \beta\kappa_{p,q}\right).
\ea
The parameter $\Omega'= \Omega / \de \omega$ is the analysis frequency relative to the OPOs cavity bandwidth for the twin beams, and $\beta=p/p_0$. The expressions for $\de p_{1,2}$ and $\de q_{1,2}$ are directly obtained from the appropriate linear combinations of the above equations. The covariance matrix is calculated as $V=\ave{\vec{x}^T\,\vec{x}}$, with $\vec{x}=(\de p_1\,, \de q_1\,, \de p_2\,, \de q_2\,, \de p_0 \,,\de q_0)$ (the superscript ``$\mathrm{out}$'' has been dropped for the sake of notational simplicity). 

\begin{figure}[ht]
\centering
\epsfig{file=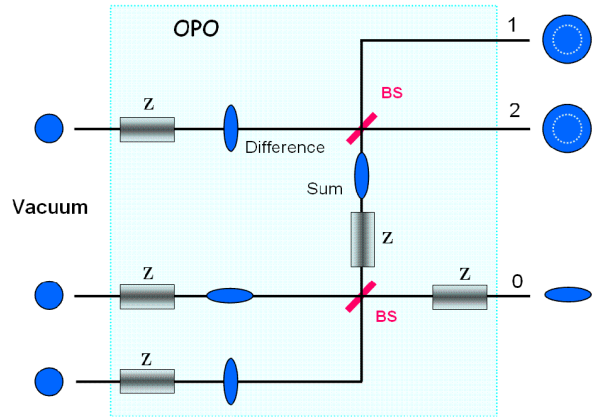,scale=0.32}
\caption{(Color online) Elementary symplectic transformations bringing a three-mode vacuum state into the fields produced by an OPO: pump (0), signal (1) and idler (2). A beam splitter transformation is represented by BS and a single-mode squeezer by Z. The circles and ellipses represent the fields' quadrature noise powers in phase space. The existence of tripartite entanglement among the beams 0, 1, and 2 is clearly seen as arising from the interference of squeezed beams in this equivalent scheme.}
\label{tripa-transf-simpletica}
\end{figure}

The main characteristics of this three-beam system were shown to be the following. The twin beams are entangled due to strong intensity correlation and phase anti-correlation~\cite{firsttwin87,prlentangtwinopo}. Because of pump depletion which always occurs above threshold, the twin beams are able to influence back the pump field, occasioning entanglement between the sum of twins and the reflected pump field. As a consequence, the OPO directly produces tripartite entangled light fields~\cite{prltrientangopo,optletttripqucorr}.


\section{Tripartite entanglement}
\label{sectripa}

The three beams generated by a single OPO form a pure quantum system, and hence can be described by a symplectic transformation acting on the three-mode vacuum field, $V_S=S_\mathrm{opo} \mathbf{1} S_\mathrm{opo}^T$, where ${\bf 1}$ is the $6\times6$ identity matrix. In order to provide an intuitive picture of the tripartite entanglement, Fig.~\ref{tripa-transf-simpletica} illustrates that the above--threshold optical parametric oscillator can be thought of as a device which realizes the symplectic transformation
\be
S_\mathrm{opo}=R^{\pi/4}_{12}\, Z\,  R^{\pi/4}_{0+}\, Z_{0+}\;,
\ee
where
\ba
R^{\theta}_{i'j'}&\hspace{-5pt}=& \hspace{-5pt}
\left( \begin{array}{cccc}
\cos\theta & 0 & \sin\theta&0 \\
0 & \cos\theta & 0& \sin\theta\\
-\sin\theta & 0 & \cos\theta&0 \\
0 & -\sin\theta & 0 & \cos\theta
\end{array} \right) ,\hspace{15pt} \\
&& \nonumber \\
Z_{i'j'}&\hspace{-5pt}=& \hspace{-5pt}\mathrm{diag}\left[e^{r_{i'}},e^{-r_{i'}},e^{r_{j'}},e^{-r_{j'}}\right] ,\hspace{15pt}
\ea
are the matrices describing the beam splitter and the squeezer transformations, respectively. The subscripts $i',j'\in\{0,1,2,+,-\}$ indicate the subspace to which the transformation is applied; absence of subscript means that it acts on the complete system. The squeezing parameter $r$ is such that the same amplitude and phase noise power occur in both modes. In addition, the OPO displaces the fields in phase space by large amounts $\A_j$; however, these operations affect only the classical mean amplitudes, and are therefore ignored in the treatment of the fields' quantum properties. 

The existence of entanglement and its structure (i.e. the way it is distributed among the fields) is revealed by the PPT criterion. We employ Eqs.~(\ref{psumout})--(\ref{qdifout}) to construct $V_S$. The partial transposition operation may then be applied to either the pump beam, resulting in the PT covariance matrix $\tilde V_S^{(0)}$, or to one of the twin beams, yielding $\tilde V_S^{(1)}=\tilde V_S^{(2)}$ (a tilde indicates partial transposition with respect to the field mode labelled by the superscript). For this type of partition, at maximum one symplectic eigenvalue $\tilde\nu^{(j)}_{k}$ from each PT matrix may assume a value smaller than one~\cite{serafini}, in this case demonstrating entanglement by violation of Eq.~(\ref{uncprinsympeigenpt}). Therefore, we always consider the \emph{smallest} symplectic eigenvalue $\tilde\nu^{(j)}$ to witness entanglement. 

We employ in our numerical results the typical experimental values $\g_0=0.05$ and $\g=0.01$ for the OPO cavity mirrors transmissions, analysis frequency $\Omega'=0.5$ (relative to cavity bandwidth) and pump power $\sigma=1.5$ (relative to threshold). The entanglement between pump and twins is quantified by $\tilde\nu^{(0)}=0.43<1$. In the same manner, each twin beam is found to be strongly entangled to the remaining two beams, since $\tilde\nu^{(1)}=\tilde\nu^{(2)}=0.20$.  These two eigenvalues are very small close to threshold ($\sigma\approx1$) and remain smaller than one even for high pump power. On the other hand, $\tilde\nu^{(0)}$ reaches one at threshold, having its minimum (maximum entanglement) around the chosen value of $\sigma=1.5$~\cite{prltrientangopo}. This is physically explained by the fact that very close to threshold all quantum correlations reside between the twin beams; on the other hand, at $\sigma=1.5$ the three output fields exhibit similar output mean power, usually the best situation for the equal sharing of correlations. Finally, all three symplectic eigenvalues are very small ($\tilde\nu \rightarrow 0$) for null analysis frequency and approach unity as $\Omega'$ increases, since pump-signal-idler quantum correlations are well known to reach their maxima at lower frequencies. These results are enough to characterize the tripartite entanglement as genuine~\cite{gklc-3modegaus}, but the structure of entanglement is further revealed by the reduced covariance matrices. 

\begin{figure}[ht]
\centering
\epsfig{file=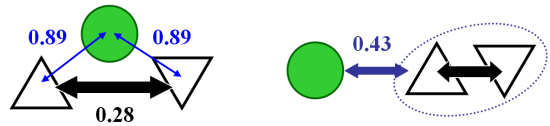,scale=0.42}
\caption{(Color online) Tripartite entanglement structure for pump, signal, and idler (circle, up-triangle, and down-triangle, respectively). The numbers represent the smallest symplectic eigenvalues of the PT matrix connecting the indicated subsystems. On the left side, entanglement between each pair of beams tracing out the third is presented. The overall structure of entanglement among the three beams is illustrated on the right, with pump entangled to the signal and idler highly entangled system.}
\label{tripa-estrutura}
\end{figure}

Tracing out the pump beam, we find that the PT reduced matrix $\tilde V_{12}$ (signal and idler) still possesses a small symplectic eigenvalue, $\tilde\nu_{\mathrm{12}}=0.28<1$. Thus, signal (idler) is strongly entangled to the remaining two beams, as shown before, because it is actually highly entangled to idler (signal). On the other hand, the PT reduced matrix for the bipartite pump and idler or pump and signal systems, $\tilde V_{01}=\tilde V_{02}$, possesses a symplectic eigenvalue close to 1, $\tilde\nu_{\mathrm{01}}=\tilde\nu_{\mathrm{02}}=0.89$, suggesting a reduced degree of entanglement between pump and a single twin. In fact, for a broad range of experimental parameters, the system presents $\tilde\nu_{0j}\lsim 1$, but, contrarily to the other symplectic eigenvalues, it tends to one as the analysis frequency $\Omega'$ is set close to zero. This result is again understood in terms of the increasing correlations between twins as $\Omega\rightarrow0$, such that $\tilde\nu_{\mathrm{12}}\rightarrow 0$, implying perfect entanglement between twins beams and no entanglement with pump in this limit.

It follows from this discussion that the pump beam is more entangled to the signal and idler composite subsystem than to either signal or idler alone (i.e. $\tilde\nu^{(0)}<\tilde\nu_{\mathrm{01}}$). One concludes that the tripartite entanglement in the OPO is not just a consequence of three pairs of bipartite entangled subsystems, since there is a finite amount of entanglement recoverable only in the complete three--beam system. Fig.~\ref{tripa-estrutura} pictorially represents the distribution of entanglement among the pump, signal, and idler beams for a single OPO. 


\section{Scalable CV entanglement}
\label{secscalable}

\begin{figure}[ht]
\centering
\epsfig{file=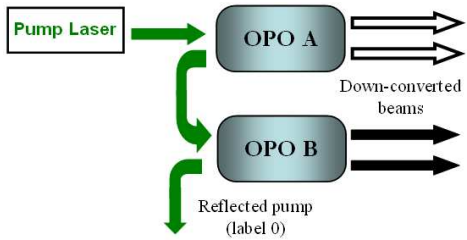,scale=0.45}
\caption{(Color online) Sketch of the system considered. Two OPOs are pumped in a cascaded configuration, generating scalable entanglement among five beams. Pump beam distributes the entanglement.}
\label{opo}
\end{figure}

We now consider a second OPO (designated ``OPO $B$'') pumped by the beam reflected from the single OPO of last section (``OPO $A$''), in a chain configuration (Fig.~\ref{opo}). The pump power reflected by the first OPO ($P_0^{\mathrm{out}\,A}$) is a decreasing function of the incident pump power ($P^\mathrm{in}$), since a larger fraction of its energy is down-converted for increasing input power. Therefore, the second OPO must have a lower threshold power ($P_\mathrm{th}^B$) than the first ($P_\mathrm{th}^A$) in order to operate above its oscillation threshold. This could be accomplished either by choosing appropriate mirror transmissions or nonlinearity strength $\chi^2$, since $P_\mathrm{th}\propto\g_0 \g^2/\chi^2$. It turns out that the ratio between the thresholds is the most relevant factor for the entanglement between the two pairs of twins, as we will show later. Therefore, as $\sigma_A\equiv P^\mathrm{in}/P_\mathrm{th}^A$ increases, $\sigma_B\equiv P_0^{\mathrm{out}\,A}/P_\mathrm{th}^B$ decreases, following the relation $\sigma_B=(\sqrt{\sigma_A} - 2 )^2 \,P_\mathrm{th}^A/P_\mathrm{th}^B$.

The pentapartite system is composed of the two pairs of twin beams generated by the two OPOs (denoted by $1_A$, $2_A$, $1_B$, and $2_B$), and the pump beam reflected by the second OPO (denoted by $0$). We define the covariance matrix $V_C=\ave{\vec{x}_C^T\,\vec{x}_C}$ for the composite system, with 
\ba
\vec{x}_C &=&(\de p_1^A,\; \de q_1^A,\; \de p_2^A,\; \de q_2^A,\; \\
& &\de p_1^B,\; \de q_1^B,\; \de p_2^B,\; \de q_2^B,\;\de p_0, \;\de q_0)\,. \nonumber
\ea

The twin fields produced by the OPO $A$ respect exactly the same solutions of Eqs.~(\ref{psumout})--(\ref{qdifout}), such that $\{\de p_s^A,\de q_s^A\}=\{\de p_s,\de q_s\}$. The solution for the OPO $B$ is found in a straightforward way. As the subtraction of twin quadratures does not depend on the input pump, $\de p_-^B$ and $\de q_-^B$ are also given by Eqs.~(\ref{pdifout}),(\ref{qdifout}). As for the sum and reflected pump subspaces, the same solutions of Eqs.~(\ref{psumout})--(\ref{q0out}) apply for $\de p_s^B$ and $\de q_s^B$ using as input pump instead the reflected pump from the OPO $A$, i.e. by the substitutions $\{\de p_0^\mathrm{in},\de q_0^\mathrm{in}\}\rightarrow\{\de p_0^A,\de q_0^A\}$. In this manner, quantum correlations from the pump beam reflected from the OPO $A$ are transferred to the twin beams produced by the OPO $B$.

In order to demonstrate genuine pentapartite entanglement, it is sufficient to show that all possible bipartitions of the system are entangled, since the complete system is pure. We use the numerical values $\g_0^B=0.04$ and $\g^B=0.0075$ for the cavity mirrors of OPO $B$ (the parameters for OPO $A$ are the same used in the last section), and $P_\mathrm{th}^B/P_\mathrm{th}^A=0.45$. We begin by studying the system behavior as a function of the incident pump power at the fixed analysis frequency $\Omega'=0.1$ (relative to the cavity bandwidth of OPO $A$). 

\begin{figure}[ht]
\centering
\epsfig{file=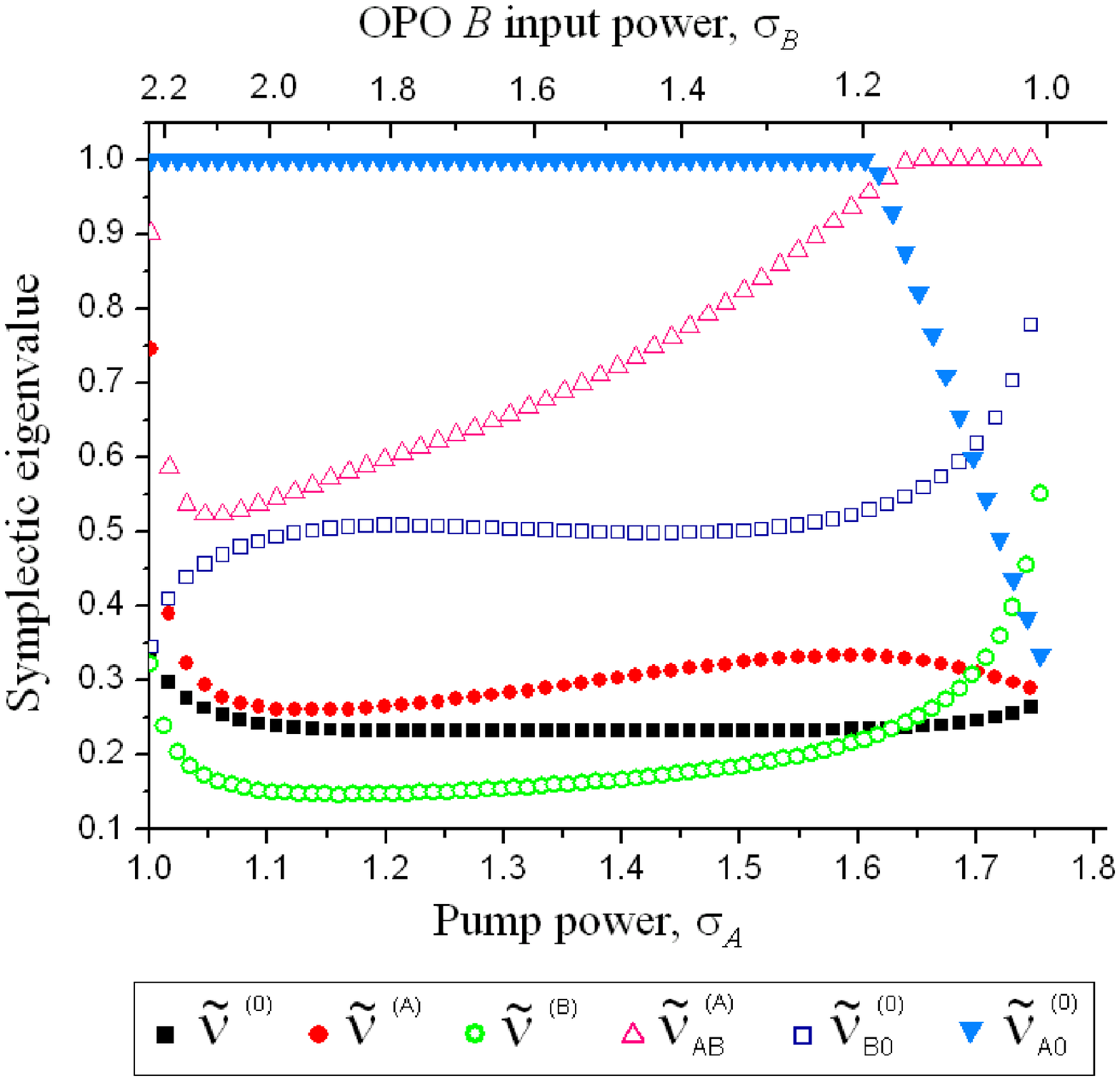,scale=0.5}
\caption{(Color online) Smallest symplectic eigenvalues obtained by applying the PT operation to various partitions of the total covariance matrix. The superscripts indicate the transposed subsystems, while subscripts, when present, indicate which reduced system is being considered. A value below one proves the entanglement between the subset of transposed subsystems and the remaining ones. }
\label{figsympeigen}
\end{figure}

First of all, we investigate how each field is entangled to the remaining ones (partitions of the form $1\times4$). The set of the \emph{smallest} symplectic eigenvalues $\tilde \nu^{(j)}$ resulting from the partial transposition $\tilde V^{(j)}$ with respect to each one of the five beams presents the following properties. The first result is that each of the four down-converted beams is highly entangled to the remaining four beams, $\tilde \nu^{(1_A)},\tilde \nu^{(2_A)},\tilde \nu^{(1_B)},\tilde \nu^{(2_B)}\ll1$. This was to be expected, since strong entanglement is actually present inside each pair of twin beams. The interesting result regards the pump beam, for which $\tilde\nu^{(0)}\ll1$ (Fig.~\ref{figsympeigen}, full squares) attests that its strong entanglement to the two pairs of twins remains. 

In the same manner, we consider partitions of the type $2\times3$. Fig.~\ref{figsympeigen} summarizes the results. Entanglement between each pair of twins and the remaining three beams is attested by the smallest symplectic eigenvalues obtained by transposing either subspace $A$ or $B$, $\tilde\nu^{(A)}<1$ (full circles) or $\tilde\nu^{(B)}<1$ (open circles). One sees that $\tilde\nu^{(0)}\rightarrow \tilde\nu^{(A)}$ when $\sigma_B \rightarrow 1$, and similarly $\tilde\nu^{(0)}\rightarrow \tilde\nu^{(B)}$ when $\sigma_A \rightarrow 1$, as expected for the limit case when just one OPO oscillates. We note that the value of $\tilde\nu^{(A)}$ at $\Omega'=0$ is the same as previously obtained by transposing the reflected pump beam from a single OPO (denoted as $\tilde\nu^{(0)}$ in last section): the twins from OPO $A$ are {\it as entangled} to the remaining subsystem of three beams as they were to the single OPO reflected pump in Sec.~\ref{sectripa}, indicating that the original entanglement between twins $A$ and reflected pump $A$ is redistributed to the three new beams. This intuitive picture is not strictly valid anymore for higher analysis frequencies, for which the original entanglement can decrease slightly. Finally, we have observed that the symplectic eigenvalues slightly decrease for lower values of $\Omega'$. In addition, their behaviors do not depend significantly on the other experimental parameters, implying a robust pentapartite entanglement.

\begin{figure}[ht]
\centering
\epsfig{file=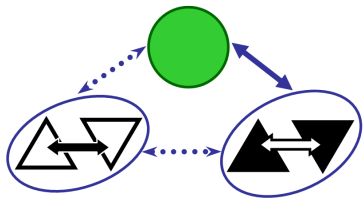,scale=0.28}
\caption{(Color online) Entanglement distribution among the five beams: twins~$A$ (open triangles), twins~$B$ (full triangles), and reflected pump (circle). The full arrows connect subgroups which violate the PPT criterion, while the dashed ones point out that the entanglement structure changes with $\Omega'$, $\sigma$, and $P_\mathrm{th}^{B}/P_\mathrm{th}^{A}$.}
\label{penta-estrutura}
\end{figure}

These results suffice to prove the existence of pentapartite entanglement in the system. It is not necessary to test partitions of the kind ``signal $A$ and pump related to the remaining three beams,'' since we know signal $A$ is highly entangled to idler $A$ and, as a consequence, to the system of remaining beams. The symmetry included in the system by the existence of two pairs of highly entangled beams reduces the number of effective partitions which need to be tested. The important results for now are the entanglement between pump and the remaining four beams, and between each pair of twins and the three remaining beams.

The entanglement structure is again better revealed by the reduced covariance matrices. The results are sketched in Fig.~\ref{penta-estrutura}. We trace out pump subspace to obtain the reduced covariance matrix $V_{AB}$ for both pairs of twins, and apply the PT operation on one pair of twins. In Fig.~\ref{figsympeigen}, the smallest symplectic eigenvalue $\tilde\nu_{\mathrm{AB}}^{(A)}<1$ (open triangles) attests the entanglement between the two pairs of twins for $\sigma_A\lsim1.65$. Therefore, the pump beam reflected by the first OPO can effectively entangle both OPOs outputs. The eigenvalue tends to one from below ($\tilde\nu_{\mathrm{AB}}^{(A)}\rightarrow1$) as $\sigma_A\rightarrow1$; for increasing pump power, it changes from $\tilde\nu_{\mathrm{AB}}^{(A)}<1$ to $\tilde\nu_{\mathrm{AB}}^{(A)}=1$, modifying the entanglement structure.

The next reduced subsystem $V_{B0}$ excludes the twins~$A$ to probe the entanglement present between pump and the twins~$B$. The partial transposition of the pump field, resulting in $\tilde\nu_{\mathrm{B0}}^{(0)}<1$ (open squares), shows that they are always entangled. Its value increases (weaker entanglement) as the threshold of OPO~$B$ is approached ($\sigma_B\rightarrow1$), which is to be expected for a single OPO (Sec.~\ref{sectripa}).

\begin{figure}[ht]
\centering
\epsfig{file=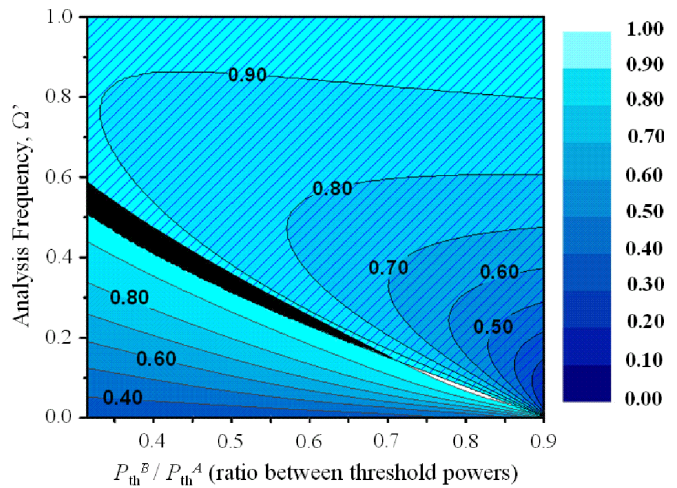,scale=0.37}
\caption{(Color online) Contour plot for the smallest symplectic eigenvalue considering the reduced subsystems: pump and twins~$A$ (patterned region) and twins~$A$ and twins~$B$ (region without pattern). The structure of pentapartite entanglement changes according to $\Omega'$ and $P_\mathrm{th}^{B}$. The small black region correspond to a situation where both $\tilde\nu_{\mathrm{AB}}^{(A)}$ and $\tilde\nu_{\mathrm{A0}}^{(0)}$ are below one, although with a weak entanglement. It has been used input pump power $\sigma_A=1.1$.}
\label{penta-contour}
\end{figure}

Finally, the reduced covariance matrix $V_{A0}$ of pump and twins~$A$ is transposed relative to pump in order to probe their entanglement. It turns out that $\tilde V_{A0}$ fulfills the uncertainty principle for lower input pump powers ($\sigma_A\lsim1.6$), since $\tilde\nu_{\mathrm{A0}}^{(0)}=1$ (full triangles), meaning that pump is not entangled to the twins~$A$ alone in this region. In view of the previous results, this indicates that the entanglement between twins $A$ and the pump beam prior to interaction with OPO $B$ is strongly converted into entanglement between the pairs of twins. At the same time, the scalability is warranted by the entanglement appearing between twins $B$ and the final reflected pump. On the other hand, for $\sigma_A\gsim1.6$ the pump remains entangled to twins $A$ (as for a single OPO), but fails to transfer these correlations to the two pairs of twins, which remain only indirectly connected to each other by the entanglement they share with the pump beam. 

Thus, as the incident pump power is varied, an alteration in the entanglement structure takes place, switching between $\tilde\nu_{\mathrm{AB}}^{(A)}<1$ and $\tilde\nu_{\mathrm{A0}}^{(0)}<1$. One may choose appropriate parameters (analysis frequency $\Omega'$ and OPO $B$'s threshold power) in order to have one or both of the entanglement types: between pump and twins~$A$ or between the two pairs of twins. Fig.~\ref{penta-contour} depicts this behavior at the fixed input pump power $\sigma_A=1.1$. The region without pattern corresponds to $\tilde\nu_\mathrm{AB}^{(A)}<1$ and the patterned region to $\tilde\nu_\mathrm{A0}^{(0)}<1$. Stronger entanglement (darker tones) between the pair of twins is found for small values of $\Omega'$ and $P_\mathrm{th}^{B}/P_\mathrm{th}^{A}$, while pump and twins $A$ get more entangled as the OPOs become similar ($P_\mathrm{th}^{B}\approx P_\mathrm{th}^{A}$). A small coexistence region, although presenting weak entanglement ($1>\tilde\nu_{\mathrm{AB}}^{(A)} \approx \tilde\nu_{\mathrm{A0}}^{(0)} \gtrsim 0.90$, seen at $\sigma_A\approx1.63$ in Fig.~\ref{figsympeigen}), is depicted by the black area. We also note that it is possible to adjust the ratio between the OPOs threshold powers either by selecting adequate crystals or cavity mirrors, since these results are not much sensitive to the exact choice of mirror transmissions. 

As a last remark, we mention that the inclusion of small spurious losses ($\lsim0.05\times \g_0,\g$) does not change the qualitative behavior of the system. The presented symplectic eigenvalues become larger, indicating a loss of quantum correlations among the beams. Intracavity losses in the twins modes have an important quantitative effect in this direction, specially concerning the entanglement between the pairs of twins. Spurious losses in the pump mode have the main consequence of decreasing the available power for pumping the OPO $B$, effectively reducing the horizontal axis range of Fig.~\ref{figsympeigen} (without altering its qualitative features). The most affected symplectic eigenvalue in this case is again related to the entanglement between the pairs of twins. In both cases, a limited region in $\sigma_A$ then appears in which neither the reduced subsystems of pairs of twins nor twins $A$ and pump are entangled. Nevertheless, full inseparability is still attested.


\section{Conclusion}
\label{secconclusion}

The entanglement existent among the three fields produced by an above-threshold optical parametric oscillator allows the implementation of a scalable network of multipartite entangled light beams, by successively employing the pump beam reflected by one OPO to pump a subsequent one. The pump beam acts as an entanglement distributor among otherwise independent OPOs. 

Moreover, the internal structure of the final pentapartite entangled state can be manipulated to some extent by tuning the incident pump power. The analysis of the reduced covariance matrices considering two cascaded OPOs reveals situations where the pump beam is either entangled to each pair of twins considered alone or both pairs are entangled to each other. In any of these cases the system is scalable, since each beam is always entangled to the remaining four beams.

The major limitation concerning the maximum number of beams which can be entangled comes from the drop in pump power as more OPOs are added to the multipartite system. However, by choosing high-quality optics (low spurious losses) and periodically poled crystals which allow a very low oscillation threshold, the number of entangled beams could be increased.

One special feature, the involvement of nonlinearities, distinguishes this system from other CV multipartite state generation based on interference, since this scheme entangles various spectral regions and, therefore, allows quantum information to be conveyed among different parts of the spectrum. Multicolor quantum networks offer the possibility to communicate quantum hardwares with otherwise incompatible working frequencies. 


\section*{Acknowledgements}

This work was supported by Funda\c{c}\~ao de Amparo \`a Pesquisa do Estado de S\~ao Paulo (FAPESP) and the European Commission through the SCALA network.



\begin{thebibliography}{99}

\bibitem{epr_physrev35}
A. Einstein, B. Podolsky, and N. Rosen,
Phys. Rev. {\bf 47}, 777 (1935).

\bibitem{bennettvicenzoreview}
C. H. Bennett and D. P. DiVincenzo, Nature {\bf 404}, 247 (2000).

\bibitem{ekertprl91}
A. K. Ekert, Phys. Rev. Lett. {\bf 67}, 661 (1991); 
M. Hillery, V. Bu\v{z}ek, and A. Berthiaume, Phys. Rev. A {\bf 59}, 1829 (1999).

\bibitem{telet-tripa-furusawa}
H. Yonezawa, T. Aokl, and A. Furusawa,
Nature, {\bf 431}, 430 (2004).

\bibitem{pengentswap}
X. Jia {\it et al.}, Phys. Rev. Lett. {\bf 93}, 250503 (2004).

\bibitem{nielsen-chuang}
M. A. Nielsen and I. L. Chuang, {\it Quantum Computation and Quantum Information} (Cambridge Univ. Press, Cambridge, 2000).

\bibitem{merminprl90}
N. D. Mermin, Phys. Rev. Lett. {\bf 65}, 1838 (1990).

\bibitem{onewaycvcomput_prl01}
R. Raussendorf and H. J. Briegel,
Phys. Rev. Lett. {\bf 86}, 5188 (2001).

\bibitem{cvclusterstatecomput_prl06}
N. C. Menicucci {\it et al.},
Phys. Rev. Lett. {\bf 97}, 110501 (2006).

\bibitem{lanceprl04}
A. M. Lance {\it et al.}, Phys. Rev. Lett. {\bf 92}, 177903 (2004).

\bibitem{weinfurterprl07}
S. Gaertner, C. Kurtsiefer, M. Bourennane, and H. Weinfurter, Phys. Rev. Lett. {\bf 98}, 020503 (2007).

\bibitem{ppt-review}
S. L. Braunstein and P. van Loock, Rev. Mod. Phys. {\bf 77}, 513 (2005);
N. J. Cerf, E. S. Polzik, and G. Leuchs, {\it Quantum Information With Continuous Variables of Atoms and Light} (Imperial College Press, 2007).

\bibitem{multiteleportnetw}
P. van Loock and S. L. Braunstein, Phys. Rev. Lett. {\bf 84}, 3482 (2000).

\bibitem{clusterpeng}
X. Su {\it et al.},
Phys. Rev. Lett. {\bf 98}, 070502 (2007).

\bibitem{pfisterpra04}
O. Pfister {\it et al.}, Phys. Rev. A {\bf 70}, 020302(R) (2004).

\bibitem{chinapra07}
Wen-ping He and Fu-li Li, Phys. Rev. A {\bf 76}, 012328 (2007).

\bibitem{prltrientangopo} 
A. S. Villar, M. Martinelli, C. Fabre, and P. Nussenzveig,
Phys. Rev. Lett. {\bf 97}, 140504 (2006). 

\bibitem{ppt}
A. Peres, Phys. Rev. Lett. {\bf 77}, 1413 (1996).

\bibitem{simon}
R. Simon, Phys. Rev. Lett. {\bf 84}, 2726 (2000).

\bibitem{symplectic-simon}
R. Simon, N. Mukunda, and B. Dutta, Phys. Rev. A {\bf 49}, 1567 (1994).

\bibitem{robertson-schroedinger} 
E. Schr\"odinger, Ber. Kgl. Akad. Wiss. Berlin {\bf 24}, 296 (1930); 
H. P. Robertson, Phys. Rev. {\bf 46}, 794 (1934).

\bibitem{adessopra04}
G. Adesso, A. Serafini, and F. Illuminati, Phys. Rev. A {\bf 70}, 022318 (2004);
A. Serafini, G. Adesso, and F. Illuminati, Phys. Rev. A {\bf 71}, 032349 (2005).

\bibitem{werner}
R. F. Werner and M. M. Wolf, Phys. Rev. Lett. {\bf 86}, 3658 (2001).

\bibitem{measure-negat}
G. Vidal and R. F. Werner, Phys. Rev. A {\bf 65}, 032314 (2002).

\bibitem{firsttwin87}
A. Heidmann {\it et al.},
Phys. Rev. Lett. {\bf 59,} 2555 (1987).

\bibitem{prlentangtwinopo} 
A. S. Villar {\it et al.},
Phys. Rev. Lett. {\bf 95,} 243603 (2005).

\bibitem{fabreqopt90}
C. Fabre {\it et al.}, Quantum Opt. {\bf 2} 159 (1990).

\bibitem{optcomm04} 
A. S. Villar, M. Martinelli, and P. Nussenzveig,
Opt. Commun. {\bf 242,} 551 (2004).

\bibitem{optletttripqucorr}
K. N. Cassemiro {\it et al.},
Opt. Lett. {\bf 32,} 695 (2007).

\bibitem{serafini}
A. Serafini, Phys. Rev. Lett. {\bf 96}, 110402 (2006).

\bibitem{gklc-3modegaus}
G. Giedke, B. Kraus, M. Lewenstein, and J. I. Cirac,
Phys. Rev. A {\bf 64}, 052303 (2001).


\end{thebibliography}
\end{document}